\documentclass[aps,twocolumn,amsmath,amssymb]{revtex4}

\usepackage[english]{babel}
\usepackage{graphicx}

\begin{document}

\newcommand{\unit}[1]{\:\mathrm{#1}}        % write units

\bibliographystyle{unsrt}
\title{A coupled quantum dot laser amplifier: Raman transitions between spin singlet and triplet states}
\author{J. M. Elzerman$^\dagger$}
\author{K. M. Weiss$^\dagger$}
\author{J. Miguel-Sanchez}
\author{A. Imamo\u{g}lu}
\affiliation{Institute of Quantum Electronics, ETH Zurich, CH-8093
Zurich, Switzerland.}
\affiliation{$^\dagger$These authors contributed equally to this work.}

\maketitle

{\bf A holy grail of photonics research is the realization of a
laser that uses a single quantum emitter as the gain
medium~\cite{Mu:1992}. Such a device would exhibit a plethora of new
features, including lasing without a well-defined
threshold~\cite{Rice:1994,Bjork:1994} and output intensity
fluctuations that remain below the shot-noise
limit~\cite{Rice:1994,Jin:1994,Briegel:1996}. While single-atom
lasers have been
demonstrated~\cite{An:1994,McKeever:2003,Dubin:2010}, compact
devices capable of continuous-wave operation require monolithic
structures involving a solid-state quantum emitter. Here, we report
the observation of steady-state laser amplification in Raman
transitions between the lowest-energy entangled spin states of a
quantum-dot molecule. Absorption and resonance fluorescence
experiments demonstrate that the singlet and triplet states have
electric-dipole coupling to a common optically excited state. Fast
spin relaxation ensures population inversion on the triplet
transition when the singlet transition is driven resonantly. By
embedding the quantum-dot molecule in a cavity of modest quality
factor, a solid-state single-emitter laser could be realized.}

\begin{figure}[b]
\includegraphics[width=81.177mm]{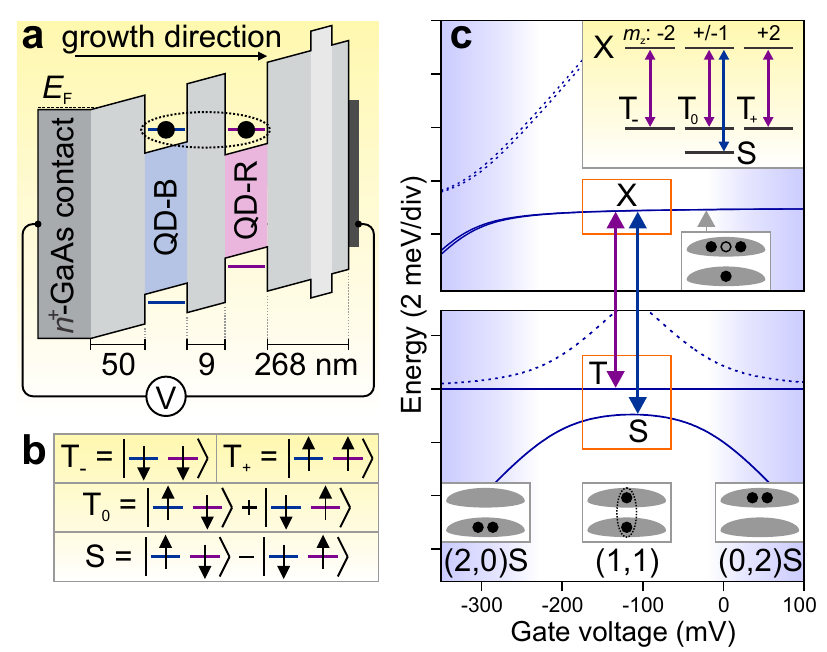}
\caption{\textbf{Coupled QD pair charged with a  single electron in
each dot. a}, Schematic energy diagram of the Schottky diode
structure with an applied gate voltage $V$. \textbf{b}, Spin states
in the (1,1) charging regime. \textbf{c}, Energy level diagram
showing the different ground states (bottom panel) and optically
excited states (top panel) versus gate voltage. State (1,1)S is
coupled via electron tunneling to states (2,0)S and (0,2)S, in which
both electrons reside in QD-B or QD-R, respectively (as illustrated
in the grey boxes, where filled circles depict electrons and open
circles depict holes). The coupling gives rise to two anticrossings
between the S states that split (1,1)S from (1,1)T, since the latter
does not experience tunnel coupling to any of the S
states~\cite{Doty:2008}. Inset: optical selection rules for
transitions from the (1,1)S and T states to the fourfold degenerate
optically excited states X. } \label{Scheme}
\end{figure}

To act as a gain medium for laser amplification, a  quantum emitter
needs to feature at least three coupled energy levels, so that
population inversion can be achieved on one transition by pumping
another ~\cite{Siegman}. A suitable level scheme is provided by a
pair of vertically stacked self-assembled InGaAs quantum dots
(QDs)~\cite{Krenner:2005,Ortner:2005,Stinaff:2006}, separated by a
thin GaAs tunnel barrier and embedded in a GaAs Schottky diode
(Fig.~\ref{Scheme}a). When both QDs contain a single electron -- a
charging regime denoted as (1,1) -- the lowest energy levels
correspond to spin singlet (S) or triplet (T$_-$, T$_0$, T$_+$)
states (Fig.~\ref{Scheme}b). Electron tunneling between the two dots
gives rise to an exchange splitting between the (1,1)S and (1,1)T
states (bottom panel in Fig.~\ref{Scheme}c), which allows us to
selectively address them optically~\cite{Gammon:2010}. The size of
the exchange splitting depends on the tunneling rate and can be
tuned by varying the gate voltage~\cite{Doty:2008}.

\begin{figure}[b]
\includegraphics[width=81.687mm]{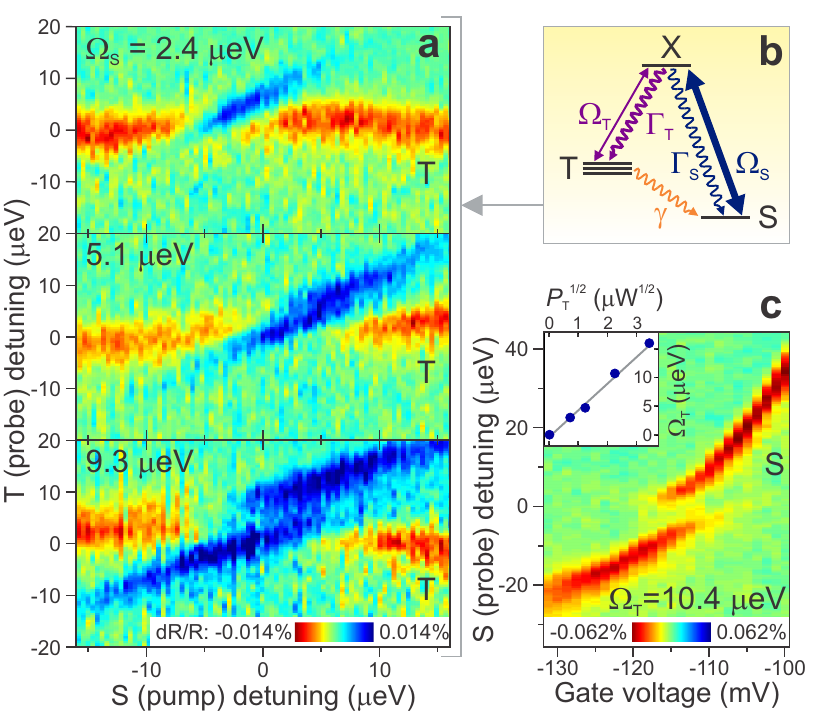}
\caption{\textbf{Laser amplification via Raman transitions. a},
Differential reflection $dR$ (in colorscale) of a weak probe laser
($\Omega_T=0.6~\unit{\mu eV}$) scanned across the T-transition, in
the presence of a pump laser (Rabi frequency $\Omega_S$) that is
stepped across the S-transition, at a fixed gate voltage of
$-94~\unit{mV}$. (The slight shift of the S-resonance energy with
increasing pump power is attributed to a laser-induced charge
buildup around the CQD.) The size of the Autler-Townes splitting
splitting in the bottom panel allows a calibration of $\Omega_S$ in
terms of the laser power on the S transition, $P_S$. \textbf{b},
Schematic energy level diagram of the lambda system formed by states
S, T \& X. $\Omega_S$ and $\Omega_T$ indicate the laser Rabi
frequencies; the effective spontaneous emission rate from X to the
combined triplet states is about three times faster than to the
singlet state ($\Gamma_T \approx 3\Gamma_S$). We observe fast
relaxation (with rate $\gamma$) from T to S. \textbf{c}, $dR$ (in
colorscale) of a weak probe laser ($\Omega_S=0.5~\unit{\mu eV}$)
scanned across the S-transition, in the presence of a strong pump
laser ($\Omega_T=10.4~\unit{\mu eV}$) resonant with the
T-transition, versus gate voltage. Inset: $\Omega_T$ as a function
of the square root of the pump laser power $\sqrt{P_T}$. $\Omega_T$
is determined from the Rabi splitting in measurements as shown in
the main panel. For identical laser powers on the S and T transition
($P_S = P_T$), we find  $\Omega_S\approx 0.9\,\Omega_T$ (see
supplementary Fig. S2 and related discussion). } \label{Ramangain}
\end{figure}

The lowest-energy optical excitation  corresponds to adding an
electron-hole pair in the top dot (QD-R), which has a redshifted
transition energy compared to the bottom dot (QD-B). The resulting
fourfold degenerate excited states X (top panel in
Fig.~\ref{Scheme}c) are labeled by the z-component of the total
angular momentum ($m_z=\pm1,\pm2$); this consists of a contribution
from the heavy hole in QD-R ($m_z=\pm\frac{3}{2}$) plus the unpaired
electron in QD-B ($m_z=\pm\frac{1}{2}$). From the associated optical
selection rules (inset to Fig.~\ref{Scheme}c), it follows that
states S and T$_0$ share two common optically excited states with
$m_z=\pm1$. At zero magnetic field, the selection rules are modified
by the hyperfine interaction with the nuclear spins, which strongly
mixes the three degenerate triplet states~\cite{Hanson:2007}.
Likewise, the four degenerate optically excited states are mixed by
both hyperfine interaction and indirect electron-hole
exchange~\cite{Doty:2008}. As a consequence, population in any X or
T level is efficiently distributed among the entire X or T manifold,
so that the full system can be represented by three levels (S, T and
X) in a simple lambda configuration, as illustrated in
Fig.~\ref{Ramangain}b. In this Letter, we use this lambda system to
achieve single-pass laser amplification of $0.014 \%$.

\begin{figure}[b]
\includegraphics[width=65.637mm]{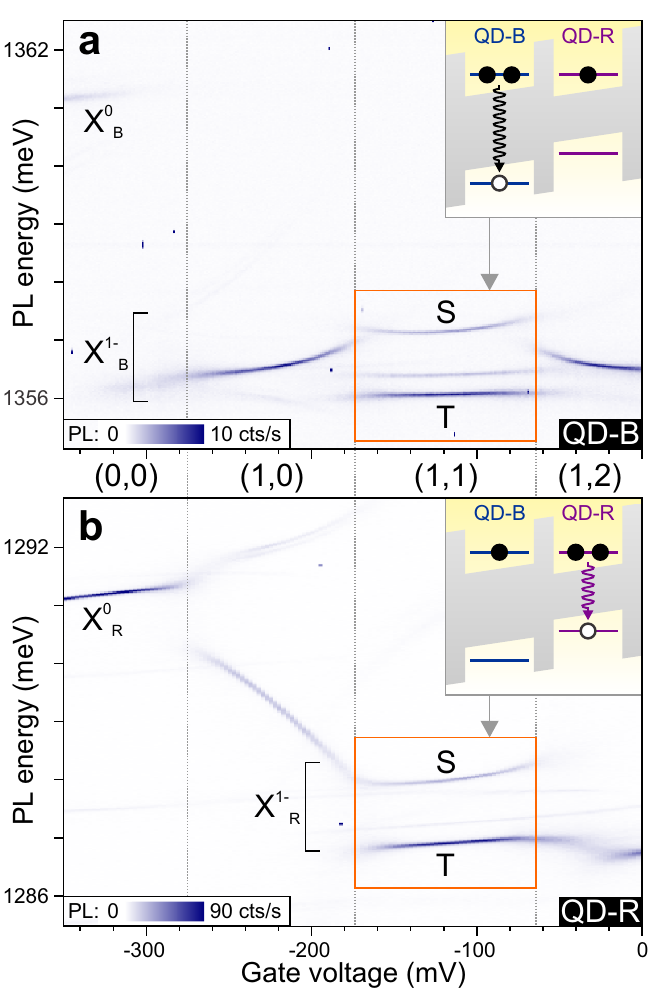}
\caption{\textbf{Identifying spin singlet and triplet transitions in a single CQD pair. a}, PL (in colorscale) from QD-B as a function of gate voltage. The dotted vertical lines separate regions with different total number of electrons in the CQD; the inferred ground state charge distribution for each region is indicated below the panel. Inside the (1,1) charging region (highlighted by the orange boxes), PL involving the S state is identified by its characteristic curvature, and by its $\sim 3$ times weaker intensity compared to PL involving the threefold degenerate T states. $X^0_B$ ($X^{1-}_B$) indicates emission from the neutral exciton (negative trion) in QD-B. Inset: schematic energy diagram illustrating $X^{1-}_B$ emission in the (1,1) regime. Because holes can tunnel from QD-B to QD-R before recombination, PL from QD-B is much weaker than that from QD-R. \textbf{b}, PL from QD-R versus gate voltage. $X^0_R$ ($X^{1-}_R$) indicates emission from the neutral exciton (negative trion) in QD-R. Inset: schematic energy diagram illustrating emission from the optically excited states X to states S or T in the (1,1) regime.
}
\label{PL}
\end{figure}

We first perform micro-photoluminescence (PL),  in order to select a
coupled quantum dot (CQD) pair that exhibits the (1,1) charging
regime. As the gate voltage is increased, the number of electrons in
the CQD increases one by one. Therefore, the PL spectra in
Fig.~\ref{PL} show typical plateaus~\cite{Warburton:2000}, separated
by dotted vertical lines indicating a change in the ground state
charge. Each plateau corresponds to emission from the neutral
exciton or negatively charged trion located in a particular dot. The
detailed shape of the plateau for a given QD depends on the number
of electrons in its partner dot, due to both tunnel
coupling~\cite{Krenner:2005,Ortner:2005,Stinaff:2006} and charge
sensing~\cite{Faelt:2008,Robledo:2008}. From these characteristic PL
patterns we identify the CQD charging sequence as
(0,0)$\rightarrow$(1,0)$\rightarrow$(1,1)$\rightarrow$(1,2). This
sequence is confirmed using numerical simulations (see supplementary
figure S1). In the (1,1) regime, we find an exchange splitting
between the S and T states of $1.1\unit{meV}$.

\begin{figure}[b]
\includegraphics[width=55.646mm]{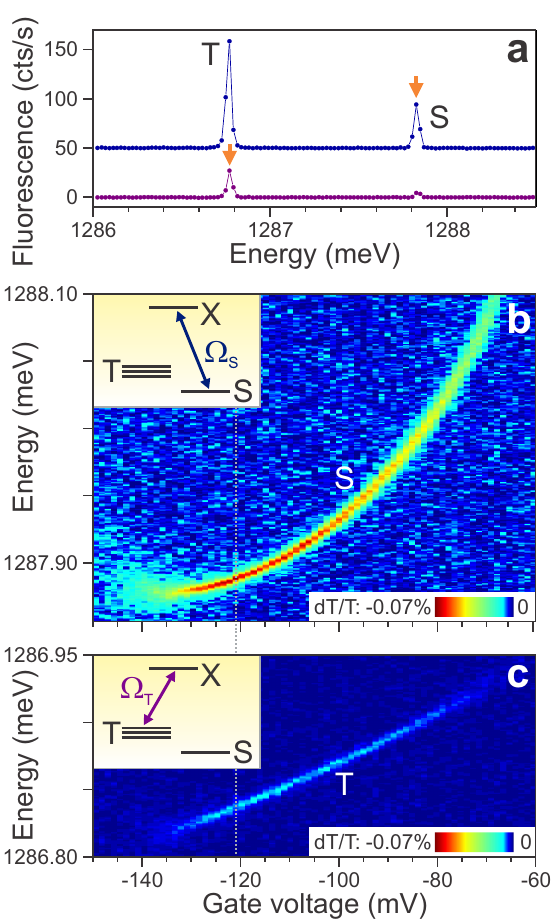}
\caption{\textbf{Characterizing the lambda system. a}, Resonance
fluorescence  detected when resonantly driving the S-X transition
(upper trace) or the T-X transition (lower trace) close to
saturation ($\Omega_{S,T}\sim 1 \unit{\mu eV}$), at a gate voltage
of $-121~\unit{mV}$. Orange arrows indicate the excitation energy.
For both traces, the T:S peak height ratio deviates slightly from 3
due to imperfect cancelation of the excitation laser. Traces have
been offset vertically for clarity. \textbf{b}, Differential
transmission $dT/T$ (in colorscale) of the S-X transition versus
gate voltage across the (1,1) regime, with $\Omega_S=0.5~\unit{\mu
eV}$. Inset: schematic energy diagram of the lambda system driven by
a laser on the S-X transition. \textbf{c}, Differential transmission
$dT/T$ (in colorscale) of the T-X transition, with
$\Omega_T=1.0~\unit{\mu eV}$; the maximum contrast is $0.011 \%$.
The linear slope of the transition versus gate voltage is due to the
DC-Stark effect (which also contributes to the slope in b). Inset:
schematic energy diagram of the lambda system driven by a laser on
the T-X transition. } \label{DT}
\end{figure}

To establish the optical connection between the S and T states,  we
employ resonance fluorescence
measurements~\cite{Vamivakas:2009,Fernandez:2009}. When resonantly
driving the S transition in QD-R (orange arrow in the upper trace of
Fig.~\ref{DT}a), fluorescence is detected not only at the same
energy (Rayleigh scattering), but also at an energy corresponding to
the T transition (Raman scattering). Conversely, when driving the T
transition in QD-R (orange arrow in the lower trace), additional
weaker emission is observed at the S transition. These measurements
demonstrate that the (1,1)S and T states indeed share common
optically excited states X in which a negative trion is located in
QD-R. Moreover, the fact that the T peak in the upper trace is $\sim
3$ times stronger than the S peak implies that the fourfold
degenerate states X are strongly mixed; without mixing, driving the
S transition would only excite the $m_z=\pm1$ subspace, resulting in
an equal number of photons emitted on the T and S transitions (see
the inset to Fig.~\ref{Scheme}c). Together, these observations
provide experimental justification for treating the system of one S,
three T and four X states as a simple lambda system, as illustrated
in Fig.~\ref{Ramangain}b.

It is important to notice that driving the T transition  results in
much less fluorescence than driving the S transition, although both
traces in Fig.~\ref{DT}a were taken with identical laser power.
Taking into account the imperfect cancelation of the excitation
laser, we find an S:T fluorescence ratio of $\sim 8$. This
surprising asymmetry is also seen in differential transmission (dT)
measurements. On the S-X transition (Fig.~\ref{DT}b), scattering of
incoming resonant laser photons leads to a maximum dT contrast of
$0.07 \%$. The dT contrast of the T-X transition (Fig.~\ref{DT}c) is
only $0.011 \%$, i.e. $\sim 6$ times lower. This difference points
towards the presence of spin relaxation from T to S. When the laser
is tuned to the S resonance, relaxation counteracts optical
shelving~\cite{Atature:2006} in the T states and thus maintains the
photon scattering rate (and therefore the dT or resonance
fluorescence signal). In contrast, a laser on the T resonance will
quickly drive the system to the S state, where it will remain
shelved for a long time, since relaxation from S back to T is
impeded by the $>1 \unit{meV}$ S-T energy difference. Thus, the
overall photon scattering rate in this case will be strongly
reduced. Using a steady-state solution of the rate equations
describing the S, T \& X populations (supplementary Fig. S2), we can
estimate the relaxation rate $\gamma$. The measured S:T scattering
ratio of $\sim 6-8$ (obtained from the difference in dT contrast
between Figs.~\ref{DT}b and \ref{DT}c or from the difference in
fluorescence intensity between the two traces in Fig.~\ref{DT}a)
gives $\gamma/\Gamma \sim 0.1-0.25$, where
$\Gamma=\Gamma_S+\Gamma_T\sim 1\unit{\mu eV}$ corresponds to the
total spontaneous emission rate from X.

Although the mechanism  behind this fast spin relaxation is not
understood at present, it is most likely related to the very large
$1.1 \unit{meV}$ exchange splitting between S and T states in our
device. (Recent experiments using CQDs with S-T splitting around
$100 \unit{\mu eV}$~\cite{Gammon:2010} did not exhibit a similar
fast relaxation.) From the perspective of using S and T$_0$ states
to encode a qubit, the spin relaxation may point to a fundamental
limitation. However, we can also use it to our advantage, since it
enables population inversion and laser amplification, as we will now
demonstrate.

To generate population inversion, we  use a strong pump laser that
is stepped across the S-X resonance. To detect the inversion, we
measure the differential reflection (dR) of a weak probe laser
scanned across the T-X resonance (see the Methods section). When the
pump is off-resonance and has modest intensity (left and right sides
of the top panel in Fig.~\ref{Ramangain}a), the probe maps out the
unperturbed T-X transition. As the pump gets closer to the S
resonance (middle of the panel), the sign of the dR signal measured
by the probe laser reverses, as indicated by the blue color. This
signifies that the probe laser actually \textit{gains} intensity by
interacting with the single CQD pair, corresponding to single-pass
laser amplification. As the pump power is increased (middle and
bottom panel in Fig.~\ref{Ramangain}a), the amplification becomes
stronger and extends further from the resonance. Remarkably, for a
pump Rabi frequency $\Omega_s = 8.0 \mu$eV  (lowest panel), we
observe amplification from the Autler-Townes doublet of dressed S
$\&$ X states \cite{Xu:2007}.

The laser amplification is directly enabled by the fast relaxation:
when the CQD decays from X to T by emitting a photon, spin
relaxation quickly depopulates the T states, and the pump field
excites the system back to X. The net result is that for
sufficiently strong relaxation rate $\gamma$ and pump rate
$\Omega_S$ (as compared to $\Gamma$), population inversion occurs on
the T-X transition. In this case, the probe laser intensity is
\textit{increased} by interaction with the CQD system. As a control
experiment, we tune the pump laser to the T transition and probe the
S transition (Fig.~\ref{Ramangain}c). In this case, a standard
(absorptive) Autler-Townes splitting is observed, without any laser
gain even for very high pump powers. This confirms that population
inversion on the S-X transition is prevented by the slow relaxation
rate from S to T at low temperatures.

In summary, we have demonstrated laser  amplification in Raman
transitions between singlet and triplet states of a single CQD
molecule. By coupling this new type of solid-state quantum emitter
to a micro-cavity, it should be possible to observe laser
oscillation. The photon statistics of such a laser are expected to
differ from ordinary lasers. From the measured $\sim 0.014 \%$
single-pass gain in probe laser intensity, we estimate that a cavity
quality factor of $\sim 7000$ should enable laser oscillation.
Employing established techniques such as solid-immersion
lenses~\cite{Vamivakas:2007} could increase the gain and thereby
reduce the required $Q$-factor by an order of magnitude, making a
practical implementation feasible.

\textbf{Methods}

\textbf{Device fabrication}

The device heterostructure was grown by molecular-beam-epitaxy on  a
(100) semi-insulating GaAs substrate. It contains two layers of
self-assembled InGaAs QDs, separated by a $9\unit{nm}$ GaAs tunnel
barrier and embedded in a GaAs Schottky diode. QDs in the top layer
tend to nucleate directly above QDs in the bottom layer due to the
strain field produced by the latter~\cite{Xie:1995}, resulting in
vertically stacked CQD pairs with density $\sim 0.1\unit{\mu
m^{-2}}$. During growth, the partially-covered-island
technique~\cite{Garcia:1998} was used to reduce the nominal
thickness of the bottom dots to $3.5\unit{nm}$, and that of the top
dots to $4.2\unit{nm}$; the resulting $\sim 50\unit{nm}$ blueshift
of the bottom dots (QD-B) with respect to the top dots (QD-R)
allowed the lowest electronic energy levels of QD-B and QD-R to be
brought into resonance. After growth, AuGe ohmic contacts were made
to the Si-doped $n^{+}$-GaAs back contact ($50\unit{nm}$ below the
bottom QD layer), and a semi-transparent gate ($2\unit{nm}$ of Ti
plus $8\unit{nm}$ of Au) was evaporated on top of the device. The
voltage $V$ applied between top gate and back contact enabled
control over the CQD charging state, allowing both QDs of a pair to
be filled with a single electron. A $40\unit{nm}$ thick
Al$_{0.3}$Ga$_{0.7}$As blocking barrier was incorporated
$10\unit{nm}$ below the top surface to reduce the current through
the device.

\textbf{Optical measurements}

The device was mounted on a 3-axis piezoelectric nanopositioning unit in a liquid-helium bath cryostat operating at $4.2 \unit{K}$. An aspheric lens with a numerical aperture of 0.55 focused the excitation light to a near-diffraction limited spot on the sample, enabling optical addressing of an individual CQD stack. For PL measurements, the CQD was excited with a non-resonant $780\unit{nm}$ diode laser; for resonance fluorescence, a narrow-band external cavity diode laser was tuned through a CQD resonance. In both cases, the resulting CQD emission was collected through the same focusing lens, coupled into a fiber and sent to a $750 \unit{mm}$ monochromator equipped with a $1200 \unit{g/mm}$ blaze-grating, which dispersed the light onto a N$_2$ cooled silicon charge-coupled device detector. To separate the CQD emission from the reflected excitation laser, both were passed through a polarizer before reaching the spectrometer, which suppressed the linearly polarized excitation laser by a factor of $\sim 10^6$.

To measure differential transmission (dT)~\cite{Alen:2003}, the intensity of a resonant laser was detected using a Si photodiode mounted directly underneath the device. To improve the signal-to-noise ratio of these measurements, a square wave with an amplitude of $100 \unit{mV}$ at a frequency of $330 \unit{Hz}$ was added to the DC gate voltage. This modulation shifted the CQD in and out of resonance with the laser via the DC Stark effect, allowing a lock-in amplifier to detect the resulting modulation amplitude of the laser light transmitted through the device.

For the two-laser differential reflection (dR) measurements, a combination of all these techniques was used. The strong pump laser and the weak probe laser had orthogonal linear polarizations; this allowed the reflected pump laser to be extinguished using a polarizer, before coupling the reflected light into a fiber and sending it to a room-temperature Si-photodiode, where it was measured using the same lock-in technique as described above.

\textbf{Acknowledgements}
We thank V. Golovach for helpful discussions, and C. Latta for help with the numerical simulations.

\end{document}